\documentstyle[12pt,epsfig]{article}
\newcommand{\tb}  {\mbox{$ \tan\beta~ $}}

\newcommand{\besg}{$b  \to  X_s \gamma~ $}

\begin{document}
\begin{flushright}
IEKP-KA/2001-19 \\[3mm]
{\tt hep-ph/0109131}\\[5mm]
\end{flushright}

 \begin{center}
  {\large\bf A global fit to the anomalous magnetic moment, \besg  
   and Higgs limits in the constrained MSSM\footnote{To appear in the Proceedings
of SUSY01, Dubna (Russia), June 2001}} \\[8mm]

  {\bf W. de Boer\footnote{Email: wim.de.boer@cern.ch}, M. Huber, C. Sander}
\\[2mm]
  {\it Institut f\"ur Experimentelle Kernphysik, University of Karlsruhe \\
       Postfach 6980, D-76128 Karlsruhe, Germany} \\[3mm]

  {\bf A.V. Gladyshev, D.I. Kazakov} \\[2mm]

{\it Bogoliubov Laboratory of Theoretical Physics,
Joint Institute for Nuclear Research, \\
141 980 Dubna, Moscow Region, Russian Federation\\[3mm]}
\end{center}

\begin{abstract}
{New data on the anomalous magnetic moment of the  muon together
with the \besg decay rate and Higgs limits are considered  within the
supergravity inspired
constrained minimal supersymmetric model. We perform a  global statistical
$\chi^2$ analysis of  these data and show that the allowed region of parameter
space is bounded from below by the Higgs limit, which depends on the trilinear
coupling and from above by the anomalous magnetic moment $a_\mu$.
}
\end{abstract}

\section{Introduction}
Recently  a new measurement of the anomalous magnetic moment of the muon became 
available, which suggests a possible 2.6 standard deviation from the Standard 
Model (SM) expectation\cite{BNL}:
$\Delta a_\mu=a_\mu^{exp}-a_\mu^{th}=(43\pm 16)\cdot 10^{-10}$.
The most popular 
explanation is given in the framework of SUSY theories\cite{we1}.
To explain the desired 
difference $\Delta a_\mu$ it requires the Higgs mixing parameter to be positive
\cite{CN} and the sparticles contributing to the chargino-sneutrino
$(\tilde{\chi}^\pm - \tilde{\nu}_\mu)$ and neutralino-smuon 
$(\tilde{\chi}^0 - \tilde{\mu})$ loop diagrams to be relatively light\cite{CM}.
The positive sign of $\mu_0$ is also preferred by the branching ratio of the 
b-quark decaying radiatively into an s-quark - \besg -\cite{we,we1}.
The error on the \besg measurements is still so large (at least 15\%),
that it does not give a significant constraint on the sparticle masses. However, it 
prefers the trilinear coupling at the GUT scale  $A_0$ to be positive for
light sparticles. In this
case of $A_0>0$ the lower limit on the Higgs mass of 114 GeV\cite{newhiggs}
becomes the most effective lower limit on the sparticle masses.
Without the \besg constraint, which implies arbitrary values of $A_0$,
the lower limit on the sparticle masses is determined by \besg.
Lack of space forces us to refer the reader to Ref.\cite{we1} for details.

\begin{figure}[htb]
\vspace*{-5mm}
\begin{center}
\epsfxsize=0.8\textwidth
\epsfbox{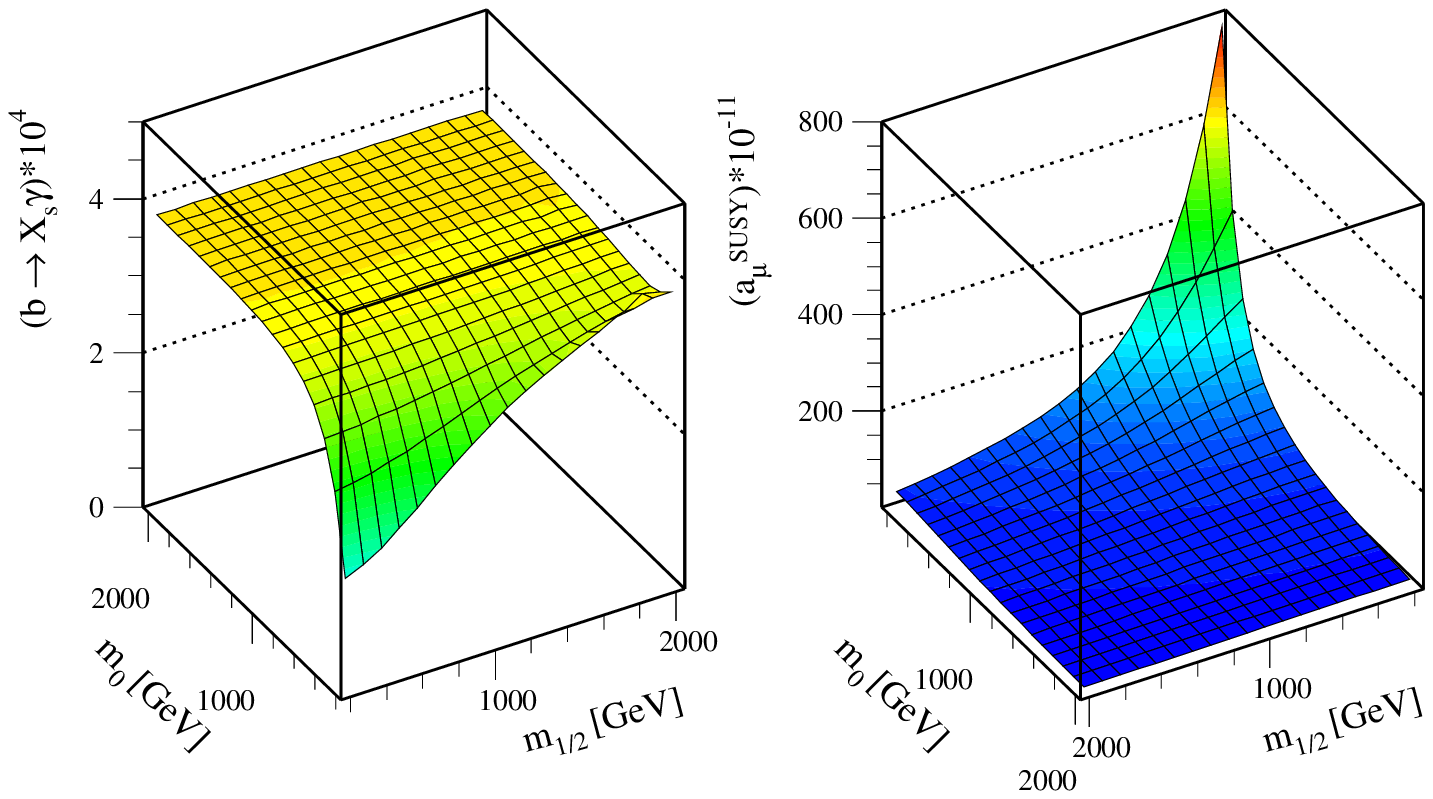}  
\epsfxsize=0.9\textwidth
\vspace*{-3mm}
\epsfbox{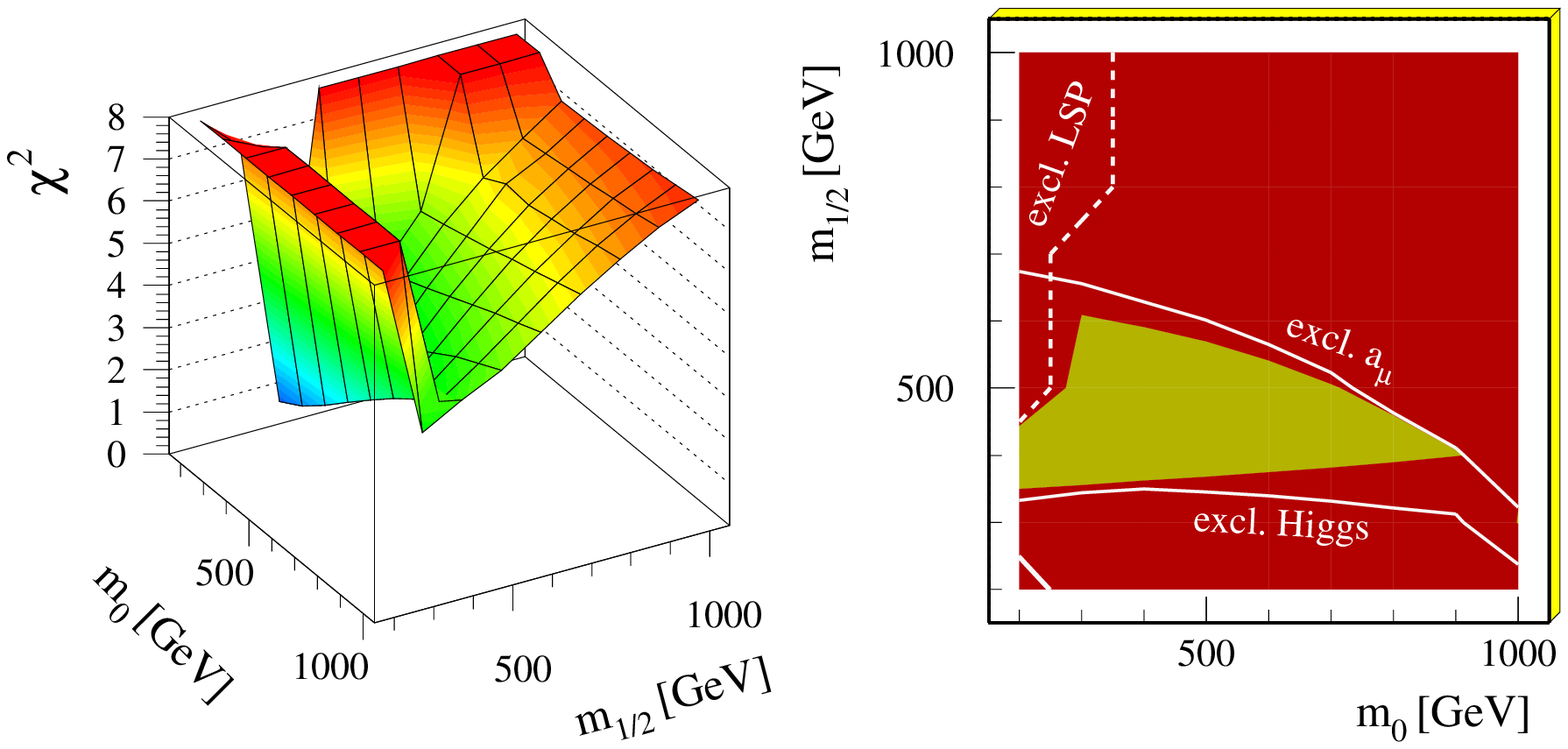}
\label{bsggm2}
\caption[]{Top: 
The values of $b\to X_s\gamma$ and $a_\mu^{SUSY}$ in the $m_0,m_{1/2}$
plane for positive $\mu$ and $\tb=35$ to be compared with experimental data 
$b\to X_s\gamma = (2.96\pm 0.46) \cdot 10^{-4}$ and 
$a_\mu^{SUSY}=(43\pm 16)\cdot 10^{-10}$.\\
Bottom:
The $\chi^2$ contribution (left) and its projection
           (right) in the $m_0, m_{1/2}$ plane for \tb= 35 and $A_0$ left free. 
           The light shaded area is the region, where the combined $\chi^2$ is below
           4. The regions outside this shaded region are excluded at 95\% C.L..
           The white lines  correspond to the  "two-sigma" contours, i.e. $\chi^2=4$ 
           for that particular contribution. The little white line at the
           left hand corner results from the \besg limit.
}
\end{center}
\end{figure}


\section{Results}

To find out the allowed regions in the parameter 
space of the CMSSM, we  fitted both the $b\to X_s\gamma$ and $a_\mu$ data 
simultaneously\cite{we1}. The fit includes the following constraints:
i) the unification of the gauge couplings, 
ii) radiative elctroweak symmetry breaking, 
iii) the masses of the third generation particles, 
iv) $b\to X_s\gamma$ and $\Delta a_\mu$, 
v) experimental limits on the SUSY masses, 
vi) the lightest superparticle (LSP) has to be neutral to be a viable candidate for 
dark matter. 
We assume common GUT scale mass parameters, i.e. $m_0$ for the spin 0 sparticles
and $m_{1/2}$ for the spin 1/2 gauginos. In addition the usual CMSSM
parameters at the GUT scale (Higgs mixing parameter  $\mu_0$, $\tb$ and trilinear
coupling $A_0$) are varied.
Since $a_\mu$ and \besg both have loop corrections with charginos their
SUSY contributions are correlated, as shown in Fig. \ref{bsggm2} (top):
the large positive SUSY contributions to $a_\mu$ for light sparticles
correspond to negative contributions for \besg.
The bottom of Fig. \ref{bsggm2} shows the combined $\chi^2$
contributions  in the $m_0$, $m_{1/2}$ plane, both in 3D and 2D.
For the 
preferred positive values of  $A_0$ the Higgs bound of 114 GeV from LEP\cite{newhiggs}
becomes an effective lower limit on $m_{1/2}$ of about 300 GeV, as shown
on the right hand bottom side of Fig. \ref{bsggm2}.
If $A_0$ is fixed at 0, the lower limit on $m_{1/2}$ is given by \besg\cite{we1}. 
These fits were made for $\tb=35$. Lower values decrease the allowed area,
for larger values the LSP limit becomes more severe\cite{we1}.

The 95\% upper limit on $m_{1/2}$ is 
determined by the lower limit on $a_\mu^{SUSY}$ and therefore depends on \tb.
For \tb=35(50) one finds $m_{1/2}\leq 610(720)$ GeV, which implies that the lightest
chargino is below 530(620) GeV and the lightest neutralino is below 270(310) GeV.
It should be noted that this upper limit strongly depends on the vacuum polarization
contributions to the fine structure constant. Recent evaluations reduce 
$\Delta a_\mu$ from a 2.6 $\sigma$ effect to less than 2 $\sigma$\cite{narison},
which increases the upper limits given above by about 25\%.
%

We thank the Heisenberg-Landau Programme, RFBR grant \# 99-02-16650
and DFG grant \# 436/RUS/113/626 for financial support.

\end{document}